\documentclass{llncs}
\usepackage{pst-node}
\usepackage{latexsym}
\usepackage{amsmath}
\usepackage{epsfig}
\usepackage{makeidx}  

\def\q5uad{\quad\quad\quad\quad\quad}

\def\mytab{\phantom{xxxx}}

\begin{document}
\mainmatter              
\title{Efficient Encoding of Watermark Numbers \\ as Reducible Permutation Graphs}




%
\titlerunning{Encoding Watermark Numbers as Graphs}  

\author{Maria~Chroni \and Stavros D. Nikolopoulos}
\authorrunning{Chroni and Nikolopoulos} 
%
\tocauthor{Maria Chroni and Stavros D. Nikolopoulos}
\institute{Department of Computer Science, University of
Ioannina, \\ GR-45110, Ioannina, Greece.\\
\email{\{mchroni,stavros\}@cs.uoi.gr}}

\maketitle              

\begin{abstract}

In a software watermarking environment, several graph theoretic watermark methods use numbers as watermark values, where some of these methods encode the watermark numbers as graph structures.
In this paper we extended the class of error correcting graphs by proposing an efficient and easily implemented codec system for encoding watermark numbers as reducible permutation flow-graphs. More precisely, we first present an efficient algorithm which encodes a watermark number $w$ as self-inverting permutation $\pi^*$ and, then, an algorithm which encodes the self-inverting permutation $\pi^*$ as a reducible permutation flow-graph $F[\pi^*]$ by exploiting domination relations on the elements of $\pi^*$ and using an efficient DAG representation of $\pi^*$. The whole encoding process takes $O(n)$ time and space, where $n$ is the binary size of the number $w$ or, equivalently, the number of elements of the permutation $\pi^*$. We also propose efficient decoding algorithms which extract the number $w$ from the reducible permutation flow-graph $F[\pi^*]$ within the same time and space complexity.
The two main components of our proposed codec system, i.e., the self-inverting permutation $\pi^*$ and the reducible permutation graph $F[\pi^*]$, incorporate important structural properties which make our system resilient to attacks.
\end{abstract}

\section{Introduction}
Software watermarking is a technique that is currently being studied to prevent or discourage software piracy and copyright infringement. The idea is similar to digital (or, media) watermarking where a unique identifier is embedded in image, audio, or video data through the introduction of errors not detectable by human perception \cite{CKLS96}.
The \emph{software watermarking problem} can be described as the problem of embedding a structure $w$ into a program $P$ such that $w$ can be reliably located and extracted from $P$ even after $P$ has been subjected to code transformations such as translation, optimization and obfuscation \cite{MC06}. More precisely, given a program $P$, a watermark $w$, and a key $k$, the software watermarking problem can be formally described by the following two functions: {\tt embed}$(P, w, k)$ $\rightarrow$ $P'$ and {\tt extract}$(P', k)$ $\rightarrow$ $w$.

Although digital watermarking has made considerable progress and become a popular technique for copyright protection of multimedia information \cite{CKLS96,TNMM04}, research on software watermarking has recently received sufficient attention. The patent by Davidson and Myhrvold \cite{DM96} presented the first published software watermarking algorithm. The preliminary concepts of software watermarking also appeared in paper \cite{G97} and patents \cite{MC96,S94}. Collberg et al. \cite{CT99,CTL98} presented detailed definitions for software watermarking. Authors of papers \cite{ZYNN03,ZTW05} have given brief surveys of software watermarking research (see, Collberg and Nagra \cite{CN2010} for an exposition of the main results).\\

\noindent {\bf Static and Dynamic Watermarking Algorithms}: There are two general categories of watermarking algorithms namely {\it static} and the {\it dynamic} algorithms \cite{CT99}. A static watermark is stored inside program code in a certain format, and it does not change during the program execution. A dynamic watermark is built during program execution, perhaps only after a particular sequence of input. It might be retrieved by analyzing the data structures built when watermarked program is running. In other cases, tracing the program execution may be required. Further discussion of static and/or dynamic watermarking issues can be found in \cite{DM96,MC96,VVS01}.\\

\noindent {\bf Algorithms and Techniques for Software Watermarking}: A lot of research
has been done on software watermarking. The major software watermarking algorithms
currently available are based on several techniques, among which the register allocation,
spread-spectrum, opaque predicate, abstract interpretation,
dynamic path techniques (see, \cite{A02,CCDHKLS04,CC04,CHC03,MIMIT00,NT04,QP98,SHKQ99}).

Recently, several software watermarking algorithms have been appeared
in the literature that encode watermarks as graph structures. In general, such encodings
make use of an encoding function {\tt encode} which converts a watermarking number $w$ into a
graph $G$, $encode(w) \rightarrow G$, and also of a decoding function {\tt decode}
that converts the graph $G$ into the number $w$, $decode(G) \rightarrow w$; we usually call
the pair $({\tt encode}, {\tt decode})$ as {\it graph codec} \cite{CKCT03}.
From a graph-theoretic point of view, we are looking for a class of graphs $\mathcal{G}$ and
a corresponding codec $({\tt encode}, {\tt decode})_\mathcal{G}$ with the following
properties:

\begin{itemize}

\item[$\bullet$\,] {\sf Appropriate Graph Types:} Graphs in $\mathcal{G}$ should be directed
having such properties, i.e., nodes with small outdegree, so that matching real program graphs;

\vspace*{0.04in}
\item[$\bullet$\,] {\sf High Resiliency:} The function $decode(G)$ should be insensitive to small
changes of $G$, i.e., insertions or deletions of a constant number of nodes or/and edges; that is,
if $G \in \mathcal{G}$ and $decode(G) \rightarrow w$ then $decode(G') \rightarrow w$ with $G' \approx G$;

\vspace*{0.04in}
\item[$\bullet$\,] {\sf Small Size:} The size $|P_w| - |P|$ of the embedded watermark should be small;

\vspace*{0.04in}
\item[$\bullet$\,] {\sf Efficient Codecs:} The functions {\tt encode} and {\tt decode} should be
computed in polynomial time.
\end{itemize}

In 1996, Davidson and Myhrvold \cite{DM96} proposed the first software watermarking algorithm which is static and embeds the watermark by reordering the basic blocks of a control flow-graph. Based on this idea, Venkatesan, Vazirani and Sinha \cite{VVS01} proposed the first graph-based software watermarking algorithm which embeds the watermark by extending a method's control flow-graph through the insertion of a directed subgraph; it is a static algorithm and is called {\tt VVS} or {\tt GTW}. In \cite{VVS01} the construction of a directed graph $G$ (or, watermark graph $G$) is not discussed. Collberg et al. \cite{CHCTS09} proposed an implementation of {\tt GTW}, which they call {\tt GTW$_{\tt sm}$}, and it is the first publicly available implementation of the algorithm {\tt GTW}. In {\tt GTW$_{\tt sm}$} the watermark is encoded as a reducible permutation graph (RPG) \cite{CKCT03}, which is a reducible control flow-graph with maximum out-degree of two, mimicking real code. Note that, for encoding integers the {\tt GTW$_{\tt sm}$} method uses only those permutations that are self-inverting.
The first dynamic watermarking algorithm ({\tt CT}) was proposed by Collberg and Thomborson \cite{CT99}; it embeds the watermark through a graph structure which is built on a heap at runtime.\\

\noindent {\bf Attacks}: A successful attack against
the watermarked program $P_w$ prevents the recognizer from extracting the watermark while
not seriously harming the performances or correctness of the program $P_w$. It is generally
assumed that the attacker has access to the algorithm used by the embedder and recognizer.
There are four main ways to attack a watermark in a software.

\begin{itemize}

\item[$\bullet$\,] {\sf Additive attacks}: Embed a new watermark into the watermarked software,
so the original copyright owners of the software cannot prove their ownership by
their original watermark inserted in the software;

\vspace*{0.04in}
\item[$\bullet$\,] {\sf Subtractive attacks}: Remove the watermark of the watermarked software
without affecting the functionality of the watermarked software;

\vspace*{0.04in}
\item[$\bullet$\,] {\sf Distortive attacks}: Modify watermark to prevent it from being extracted by
the copyright owners and still keep the usability of the software;

\vspace*{0.04in}
\item[$\bullet$\,] {\sf Recognition attacks}: Modify or disable the watermark detector, or its inputs,
so that it gives a misleading result. For example, an adversary may assert that
``his" watermark detector is the one that should be used to prove ownership in
a courtroom test.
\end{itemize}

Attacks against graph-based software watermarking algorithms can mainly occur in the following three ways:
(i) {\sf Edge-flip attacks}, (ii) {\sf Edges-addition/deletion attacks}, and (iii) {\sf Node-addition/deletion attacks}.\\

\noindent {\bf Our Contribution}:  In this paper we present an efficient and easily implemented algorithm for encoding numbers as reducible permutation flow-graphs through the use of self-inverting permutations (or, for short, SIP).

More precisely, we first present an efficient algorithm which encodes a number (integer) $w$ as self-inverting permutation $\pi^*$. Our algorithm, which we call {\tt Encode\_W-to-SIP}, takes as input an integer $w$, computes first its binary representation $b_1b_2 \cdots b_n$, then constructs a bitonic permutation on $2n+1$ numbers, and finally produces a self-inverting permutation $\pi^*$ of length $2n+1$ in $O(n)$ time and space. We also present a decode algorithm which extracts the integer $w$ from the self-inverting permutation $\pi^*$ within the same time and space complexity; we call the decode algorithm {\tt Decode\_SIP-to-W}.

Having designed an efficient method for encoding integers as self-inverting permutations, we next describe an algorithm for encoding a self-inverting permutation into a directed graph structure having properties capable to match real program graphs. In particular, we propose the algorithm {\tt Encode\_SIP-to-RPG} which encodes the self-inverting permutation $\pi^*$ as a reducible permutation flow-graph $F[\pi^*]$ by exploiting domination relations on the elements of $\pi^*$ and using an efficient DAG representation of $\pi^*$. The whole encoding process takes $O(n)$ time and requires $O(n)$ space, where $n$ is the length of the permutation $\pi^*$. We also propose an efficient and easily implemented algorithm, the algorithm {\tt Decode\_RPG-to-SIP}, which extract the self-inverting permutation $\pi^*$ from the reducible permutation flow-graph $F[\pi^*]$ by converting first the graph $F[\pi^*]$ into a directed tree $T[\pi^*]$ and then applying DFS-search on $T[\pi^*]$. The decoding process takes time and space linear in the size of the flow-graph $F[\pi^*]$, that is, the algorithm {\tt Decode\_RPG-to-SIP} takes $O(n)$ time and space. We point out that the only operations used by the decoding algorithm are edge modifications on $F[\pi^*]$ and DFS-search on trees.

It is worth noting that our codec $({\tt encode}, {\tt decode})_{F[\pi^*]}$ system incorporates several important properties which characterize it as an efficient and easily implemented software watermarking component. In particular, the reducible permutation flow-graph $F[\pi^*]$ does not differ from the graph data structures built by real programs since its maximum outdegree does not exceed two and it has a unique root node so the program can reach other nodes from the root node. The function {\tt Decode\_RPG-to-SIP} is high insensitive to small
edge-changes and quite insensitive to small node-changes of $F[\pi^*]$, and the graph $F[\pi^*]$ unable us to correct such edge changes. Moreover, the self-inverting permutation $\pi^*$ captures important structural properties, due to the bitonic property used in the construction of $\pi^*$, which make our codec system resilient to attacks.

Finally, we point out that our codec $({\tt encode}, {\tt decode})_{F[\pi^*]}$ system has very low time and space complexity which is $O(n)$ where $n$ is the number of bits in a binary representation of the watermark integer $w$. Indeed, both functions {\tt Encode\_W-to-SIP} and {\tt Decode\_SIP-to-W} are computed in time and space linear in the binary size of the watermark integer $w$. Moreover, the functions {\tt Encode\_SIP-to-RPG} and {\tt Decode\_RPG-to-SIP} are also computed in linear time and space; in particular, the function {\tt Encode\_SIP-to-RPG} is computed in time and space linear in the length of the self-inverting permutation $\pi^*$ which is $O(n)$, while the function {\tt Decode\_RPG-to-SIP} is computed in time and space linear in the size of the flow-graph $F[\pi^*]$ which is also $O(n)$.\\


\section{Preliminaries}

We consider finite graphs with no multiple edges.
For a graph~$G$, we denote by $V(G)$ and $E(G)$
the vertex set and edge set of $G$, respectively.
The \emph{neighborhood}~$N(x)$ of a vertex~$x$ of the graph~$G$ is
the set of all the vertices of $G$ which are adjacent to $x$. The
\emph{degree} of a vertex~$x$ in the graph~$G$, denoted $deg(x)$, is
the number of edges incident on $x$; thus, $d(x) = |N(x)|$. For a node $x$ of a directed graph $G$, the number of head-endpoints of the directed edges adjacent to $x$ is called the indegree of the node $x$, denoted $indeg(x)$, and the number of tail-endpoints is its outdegree, denoted $outdeg(x)$.

A \emph{path} in a graph~$G$ of length $k$ is a sequence of vertices $(v_0, v_1,
\ldots, v_k)$ such that $(v_{i-1}, v_i) \in E(G)$ for $i =
1,2,\ldots,k$.  A path is called \emph{simple} if none of its
vertices occurs more than once. A path (simple path) $(v_0, v_1,
\ldots, v_k)$ is a \emph{cycle} (\emph{simple cycle}) of length $k+1$ if $(v_0, v_k)
\in E(G)$.


Next, we introduce some definitions that are key-objects in our
algorithms for encoding numbers as graphs. Let $\pi$ be a permutation
over the set $N_n = \{1, 2, \ldots, n\}$. We think of permutation $\pi$
as a sequence $(\pi_1, \pi_2, \ldots, \pi_n)$, so, for example, the
permutation $\pi = (1, 4, 2, 7, 5, 3, 6)$ has $\pi_1 = 1$, $\pi_2 = 4$, ect.
Notice that $\pi^{-1}_i$ is the position in the sequence
of the number $i$; in our example, $\pi_4^{-1} = 2$, $\pi_7^{-1} = 4$, $\pi_3^{-1} = 6$, ect \cite{Gol80}. \\

\noindent {\bf Definition~1}: The inverse of a permutation $(\pi_1, \pi_2, \ldots, \pi_n)$ is the
permutation $(q_1, q_2, \ldots, q_n)$ with $q_{\pi_i} = \pi_{q_i} = i$.
A {\it self-inverting permutation} (or, involution) is a permutation that is its
own inverse: $\pi_{\pi_i} = i$.\\

By definition, every permutation has a unique inverse, and the inverse of the
inverse is the original permutation. Clearly, a permutation is a self-inverting
permutation if and only if all its cycles are of length 1 or 2; hereafter, we
shall denote a 2-cycle as $c=(x,y)$ and an 1-cycle as $c=(x)$, or, equivalently, $c=(x, x)$.\\

\noindent {\bf Definition~2}: Let $C_{1,2} = \{c_1 = (x_1, y_1), c_2 = (x_2, y_2), \ldots, c_k = (x_k, y_k)\}$ be the set of all the cycles of a self-inverting permutation $\pi$ such that $x_i < y_i$ $(1 \leq i \leq k)$, and let $\prec$ be a linear order on $C_{1,2}$ such that $c_i \prec c_j$ if $x_i < x_j$, $1 \leq i, j \leq k$. A sequence $C = (c_1, c_2, \ldots, c_k)$ of all the cycles of a self-inverting
permutation $\pi$ is called {\it increasing cycle representation} of $\pi$ if $c_1 \prec c_2 \prec \cdots \prec c_k$. The cycle $c_1$ is the minimum element of the sequence $C$.\\

Let $\pi$ be a permutation on $N = \{1, 2, \ldots, n\}$. We say that an element $i$ of the permutation $\pi$ {\it dominates} the element $j$ if $i > j$ and $\pi^{-1}_i < \pi^{-1}_j$. An element $i$ {\it directly dominates} (or, for short, didominates) the element $j$ if $i$ dominates $j$ and there exists no element $k$ in $\pi$ such that $i$ dominates $k$ and $k$ dominates $j$ \cite{N02}. For example, in the permutation $\pi = (8, 3, 2, 7, 1, 9, 6, 5, 4)$, the element $7$ dominates the elements $1, 6, 5, 4$ and it directly dominates the elements $1, 6$.\\

\noindent {\bf Definition~3}: The domination (resp. didomination) set dom($i$) (resp. didom($i$)) of the element $i$ of a permutation $\pi$ is the set of all the elements of $\pi$ that dominate (resp. didominate) the element $i$.\\

\noindent {\bf Definition~4}: An undirected graph $G$ with vertices numbered from 1 to $n$; that is, $V(G) = \{1, 2, \ldots, n\}$, is called a {\it permutation graph} if there exists a permutation $\pi = (\pi_1, \pi_2, \ldots, \pi_n)$ on $N_n$ such that,
$(i, j) \in E(G)$ if and only if $(i-j)(\pi_i^{-1} - \pi_j^{-1})<0$.\\


A flow-graph is a directed graph $F$ with an initial node $s$ from which all other nodes are reachable. A directed graph $G$ is strongly connected when there is a path $x \rightarrow y$ for all nodes $x$, $y$ in $V(G)$. A node $u$ is an {\it entry} for a subgraph $H$ of the graph $G$ when there is a path $p = (y_1, y_2, \ldots, y_k, x)$ such that $p \cap H = \{x\}$.\\

\noindent {\bf Definition~5}: A flow-graph is reducible when it does not have a strongly connected subgraph with two (or more) entries.
%


\section{Encode Watermark Numbers as Self-inverting Permutations}

In this section, we first introduce the notion of {\it Bitonic Permutations} and then we present two algorithms, namely {\tt Encode\_W-to-SIP} and {\tt Decode\_SIP-to-W}, for encoding an integer $w$ into an self-inverting permutation $\pi^*$ and extracting it from $\pi^*$. Both algorithms run in $O(n)$ time, where $n$ is the length of the binary representation of the integer $w$ \cite{CN10}.

\vskip 0.3in 
\subsection{Bitonic Permutations}

The key-object in our algorithm for encoding integers as self-inverting
permutations is the bitonic permutation: a permutation $\pi = (\pi_1, \pi_2, \ldots , \pi_n)$
over the set $N_n$ is called bitonic if either monotonically increases and then
monotonically decreases, or else monotonically decreases and then monotonically
increases. For example, the permutations $\pi_1 = (1, 4, 6, 7, 5, 3, 2)$ and
$\pi_2 = (6, 4, 3, 1, 2, 5, 7)$ are both bitonic \cite{CN10}.

In this paper, we consider only bitonic permutations that monotonically increases
and then monotonically decreases. Let $\pi = (\pi_1, \pi_2, \ldots, \pi_i, \pi_{i+1}, \ldots, \pi_n)$ be such a
bitonic permutation over the set $N_n$ and let $\pi_i$, $\pi_{i+1}$ be the two consecutive
elements of $\pi$ such that $\pi_i > \pi_{i+1}$. Then, the sequence $X = (\pi_1, \pi_2, \ldots, \pi_i)$ is
called first increasing subsequence of $\pi$ and the sequence $Y = (\pi_{i+1}, \pi_{i+2}, \ldots, \pi_n)$
is called first decreasing subsequence of $\pi$.

We next give some notations and terminology we shall use throughout the paper.
Let $w$ be an integer number. We denote by $B = b_1b_2\cdots b_n$ the binary representation
of $w$. If $B_1 = b_1b_2\cdots b_n$ and $B_2 = d_1d_2 \cdots d_m$ be
two binary numbers, then the number $B_1||B_2$ is the binary number $b_1b_2\cdots b_nd_1d_2 \cdots d_m$. The binary sequence
of the number $B = b_1b_2 \cdots b_n$ is the sequence $B^* = (b_1, b_2, \ldots, b_n)$ of length $n$.

Let $B = b_1b_2 \cdots bn$ be a binary number. Then, $flip(B) = b'_1b'_2 \cdots b'_n$ is the binary number
such that $b'_i = 0$ (1 resp.) if and only if $b_i = 1$ (0 resp.), $1 \leq  i \leq  n$.

\vskip 0.3in 
\subsection{Algorithm Encode\_W-to-SIP}

In this section, we present an algorithm for encoding an integer as self-inverting permutation. In particular, our algorithm takes as input an integer $w$, computes the binary representation $b_1 b_2 \cdots b_n$ of $w$, and then produces a self-inverting permutation $\pi^*$ in $O(n)$ time. We next describe the proposed algorithm:


\vspace*{0.15in} \noindent Algorithm {\tt Encode\_W-to-SIP}
\vspace*{-0.04in}
\begin{enumerate}
\item[1.\,]   
Compute the binary representation $B = b_1b_2 \cdots b_n$ of $w$;
\vspace*{0.04in}
\item[2.\,]   
Construct the binary number $B' = 00 \cdots 0||B||1$ of length $2n+1$, and then the binary sequence $B^* = (b_1, b_2, \ldots, b_{n'})$ of $flip(B')$;
\vspace*{0.04in}
\item[3.\,]   
Find the sequence $X = (x_1, x_2, \ldots, x_k)$ of the $0's$ positions and the sequence $Y = (y_1, y_2, \ldots, y_m)$ of the $1's$ positions in $B^*$ from left-to-right;
\vspace*{0.04in}
\item[4.\,]   
Construct the bitonic permutation $\pi^b = X || Y^R$ on $n' = 2n+1$ numbers;\\
let $\pi^b = (x_1, x_2, \ldots, x_k, y_m, y_{m-1}, \ldots, y_1)$
\vspace*{0.04in}
\item[5.\,]   
Set $(\pi_1, \pi_2, \ldots, \pi_k, \pi_{k+1}, \pi_{k+2}, \ldots, \pi_{n'}) = (x_1, x_2, \ldots, x_k, y_m, y_{m-1}, \ldots, y_1)$, $i=1$ and $j=n'$; \vspace*{0.04in}\\
while $i<j$ do the following:\\
\mytab construct the 2-cycle $c_i = (\pi_i, \pi_j)$, and set $i=i+1$ and $j=j-1$;\\
end-while;\\
if $i=j$ then construct the 1-cycle $c_i = (\pi_i)$;
\vspace*{0.04in}
\item[6.\,]   
Construct the permutation $\pi^* = (\pi_1, \pi_2, \ldots, \pi_{n'})$ on $n' = 2n+1$ numbers such that $\pi_i=i$, $1 \leq i \leq n'$;
\vspace*{0.04in}
\item[7.\,]   
Let $C$ be the set of all cycles computed at step 5;\\
for each 2-cycle $(\pi_i, \pi_j) \in C$ set $\pi_{\pi_i} = \pi_j$ and $\pi_{\pi_j} = \pi_i$;
\vspace*{0.04in}
\item[8.\,]   
Return the self-inverting permutation $\pi^*$;
\end{enumerate}

\vspace*{0.1in}
\noindent {\bf Example~1}: Let $w = 12$ be the input watermark integer in the algorithm {\tt Encode\_W-to-SIP}. We first compute the binary representation $B = 1100$ of the number $12$; then we construct the binary number $B' = 000011001$ and the binary sequence $B^* = (1, 1, 1, 1, 0, 0, 1, 1, 0)$ of $flip(B')$; we compute the sequences $X = (5, 6, 9)$ and $Y = (1, 2, 3, 4, 7, 8)$, and then construct the bitonic permutation $\pi = (5, 6, 9, 8, 7, 4, 3, 2, 1)$ on $n'=9$ numbers; since $n'=9$ odd, we select $4$ pairs $(5, 1)$, $(6, 2)$, $(9, 3)$, $(8, 4)$ and the number $7$ and then construct the self-inverting permutation $\pi^* = (5, 6, 9, 8, 1, 2, 7, 4, 3)$.\\

\noindent {\it Time and Space Complexity.} The encoding algorithm {\tt Encode\_W-to-SIP} performs basic operations on sequences of lengths $O(n)$, where $n$ is the number of bits in the binary representation of $w$ (see Figure~1); hereafter, for the number $n$ we shall call the term {\it binary size} of the integer $w$. Moreover, all the operations are executed in place, i.e., the algorithm uses no additional space except of a constant number of variables. It is easy to see that the whole encoding process requires $O(n)$ time and space. Thus, the following theorem holds:

\vspace*{0.04in}
\begin{theorem}\label{theo1.1}
Let $w$ be an integer and let $b_1b_2\cdots b_n$ be the binary representation of $w$. The algorithm {\tt Encode\_W-to-SIP} encodes the number $w$ in a self-inverting permutation $\pi^*$ of length $2n+1$ in $O(n)$ time and space.
\end{theorem}

\vskip 0.3in 
\subsection{Algorithm Decode\_SIP-to-W}

Next, we present an extraction algorithm, that is, an algorithm for decoding a self-inverting permutation. More precisely, our extraction algorithm, which we call Decode\_SIP-to-W, takes as input a self-inverting permutation $\pi^*$ produced by Algorithm Encode\_W-to-SIP and returns its corresponding integer $w$. The time complexity of the decode algorithm is also $O(n)$, where $n$ is the length of the permutation $\pi^*$. We next describe the proposed algorithm:


\vspace*{0.15in} \noindent Algorithm {\tt Decode\_SIP-to-W}
\vspace*{-0.04in}
\begin{enumerate}
\item[1.\,]   
Compute the increasing cycle representation $C = (c_1, c_2, \ldots, c_k)$ of the self-inverting permutation
$\pi^* = (\pi_1, \pi_2, \ldots, \pi_{n'})$, where $n' = 2n+1$, that is, $c_1 \prec c_2 \prec \cdots \prec c_k$;
\vspace*{0.04in}
\item[2.\,]   
Set $i = 1$ and $j = n'$:
\vspace*{0.04in}
\item[3.\,]   
Construct the permutation $\pi^b$ of length $n'$ as follows: \vspace*{0.04in}\\
\mytab while the set $C$ is not empty, do the following: \vspace*{0.04in}\\
\mytab\mytab Select the minimum element $c$ of the sequence $C$;\\
\mytab\mytab {\bf Case 1:} \, the selected cycle $c$ has length $2$ and let $c = (a, b)$:\\
\mytab\mytab\mytab\mytab \ $\pi_i = b$ and $\pi_j = a$;\\
\mytab\mytab\mytab\mytab \ $i = i+1$ and $j = j -1$; \vspace*{0.04in}\\
\mytab\mytab {\bf Case 2:} \,  the selected cycle $c$ has length $1$ and let $c = (a)$: \\
\mytab\mytab\mytab\mytab \ $\pi_i = a$ and $i = i+1$;\\
\mytab\mytab Remove the cycle $c$ from $C$;
\vspace*{0.04in}
\item[4.\,]   
Find the first increasing subsequence $X = (\pi_1, \pi_2, \ldots, \pi_k)$ and then the decreasing subsequence $Y = (\pi_{k+1}, \pi_{k+2}, \ldots, \pi_{k'})$ of $\pi$;
\vspace*{0.04in}
\item[5.\,]   
Construct the binary sequence $B^* = (b_1, b_2, \ldots, b_{n'})$ as follows:\\
set 0's in positions $\pi_1, \pi_2, \ldots, \pi_k$ and 1's in positions $\pi_{k+1}, \pi_{k+2}, \ldots, \pi_{k'}$;
\vspace*{0.04in}
\item[6.\,]   
Compute $B' = flip(B^*) = (b_1, b_2, \ldots, b_n, b_{n+1}, \ldots, b_{n'-1}, b_{n'})$;
\vspace*{0.04in}
\item[7.\,]   
Return the integer $w$ of the binary number $B = b_{n+1}b_{n+2} \cdots b_{n'-1}$;
\end{enumerate}

\begin{figure}[t]
 \hrule \medskip
  \centering
  \includegraphics[scale=0.7]{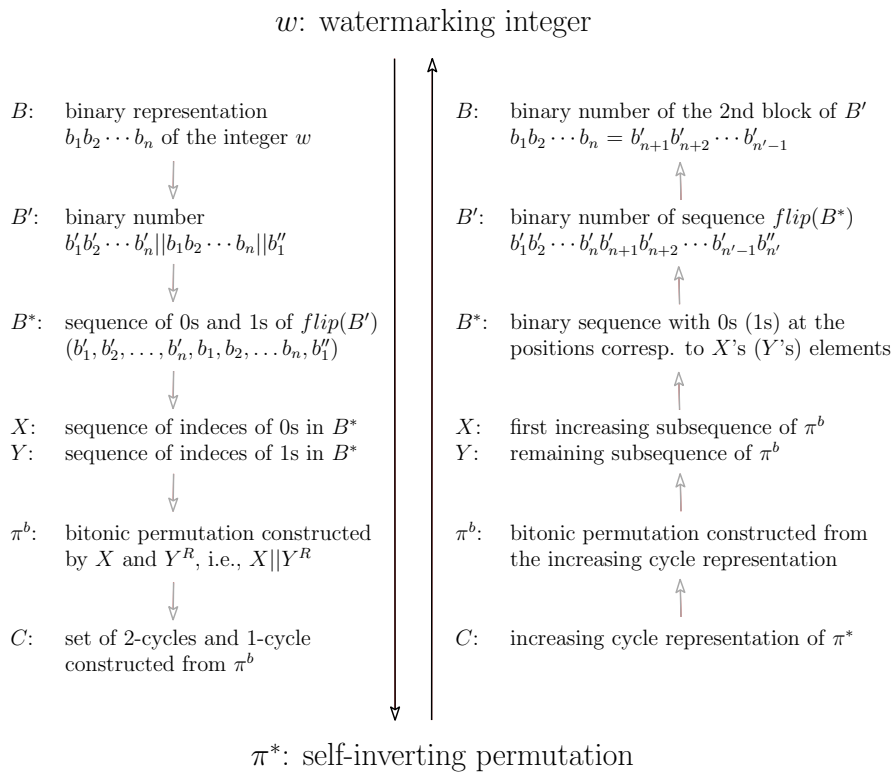}
  \centering
  \caption{\small{The main data components used by the algorithms {\tt Encode\_W-to-SIP} and {\tt Decode\_SIP-to-W}}}
  \label{class} \medskip
 \hrule
\end{figure}

\vspace*{0.04in}
\noindent {\bf Example~2}: Let $\pi^* = (5, 6, 9, 8, 1, 2, 7, 4, 3)$ be a self-inverting permutation produced by the algorithm {\tt Encode\_W-to-SIP}. The cycle representation of $\pi^*$ is the following: $(1, 5)$, $(2, 6)$, $(3, 9)$, $(4, 8)$, (7); from the cycles we construct the permutation $\pi = (5, 6, 9, 8, 7, 4, 3, 2, 1)$; then, we compute first increasing subsequence $X = (5, 6, 9)$ and the first decreasing subsequence $Y = (8, 7, 4, 3, 2, 1)$; we then construct the binary sequence $B^* = (1, 1, 1, 1, 0, 0, 1, 1, 0)$ of length $9$; we flip the elements of $B^*$ and construct the sequence $B' = (0, 0, 0, 0, 1, 1, 0, 0, 1)$; the binary number 1100 is the integer $w = 12$.\\

\noindent {\it Time and Space Complexity.} It is easy to see that the decoding algorithm {\tt Decode\_SIP-to-W} performs the same basic operations on sequences of lengths $O(n)$ as the encoding algorithm (see Figure~1). Thus, we obtain the following result:

\vspace*{0.04in}
\begin{theorem}\label{theo1.2}
Let $\pi^*$ be a self-inverting permutation of length $n$ which encodes an integer $w$ using the algorithm {\tt Encode\_W-to-SIP}. The algorithm {\tt Decode\_SIP-to-W} correctly decodes the permutation $\pi^*$ in $O(n)$ time and space.
\end{theorem}

\section{Encode Self-inverting Permutations as Reducible Permutation Graphs}
Having proposed an efficient method for encoding integers as self-inverting permutations, we next describe an algorithm for encoding a self-inverting permutation $\pi^*$ into a directed graph $F[\pi^*]$. We also describe a decoding algorithm for extracting the permutation $\pi^*$ from the graph $F[\pi^*]$.

\vskip 0.3in 
\subsection{Algorithm Encode\_SIP-to-RPG}

We next propose the algorithm {\tt Encode\_SIP-to-RPG} which takes as input the self-inverting permutation $\pi^*$ produced by the algorithm {\tt Encode\_W-to-SIP} and constructs a reducible permutation flow-graph $F[\pi^*]$ by using an efficient DAG representation of the self-inverting permutation $\pi^*$. The whole encoding process takes $O(n)$ time and requires $O(n)$ space, where $n$ is the length of the input permutation $\pi^*$.

Given a self-inverting permutation $\pi^*$ of length $n$ our decoding algorithm works on two phases:
\vspace*{-0.06in}
\begin{itemize}
\item[I.\,]
it first uses a strategy to transform the permutation $\pi^*$ into a directed acyclic graph $D[\pi^*]$ using certain combinatorial properties of the elements of $\pi^*$;
\vspace*{0.04in}
\item[II.\,]
then, it constructs a directed graph $F[\pi^*]$ on $n+2$ nodes using the adjacency relation of the nodes of the dag $D[\pi^*]$.
\end{itemize}

Next, we first describe the main ideas behind the two phases and then we present in details the whole algorithm.\\

\noindent {\bf Construction of the DAG $D[\pi^*]$ from the permutation $\pi^*$}: We construct the directed acyclic graph $D[\pi^*]$ by exploiting the didomination relation of the elements of $\pi^*$, as follows:
\begin{itemize}
\item[(i)\,]
for every element $i$ of $\pi^*$, create a vertex $v_i$ and add it to the vertex set $V(D[\pi^*])$;
\vspace*{0.04in}
\item[(ii)\,]
compute the didomination relation of each element $i$ of $\pi^*$; recall that the didomination set $didom(i)$ of the element $i$ contains all the elements $j$ of $\pi^*$ that are didominated by the element $i$ (see Definition~3);
\vspace*{0.04in}
\item[(iii)\,]
for every pair of vertices $(v_i, v_j)$ of the set $V(D[\pi^*])$ do the following: add the edge $(v_i, v_j)$ in $E(D[\pi^*])$ if the element $i$ didominates the element $j$ in $\pi^*$;
\vspace*{0.04in}
\item[(iv)\,] create two dummy vertices $s$ and $t$ and add both in $V(D[\pi^*])$; then, add the edge $(s, v_i)$ in $E(D[\pi^*])$, for every $v_i$ with $indeg(v_i)=0$, and the edge $(v_i, t)$ in $E(D[\pi^*])$, for every $v_i$ with $outdeg(v_i) = 0$.
\end{itemize}

\noindent {\bf Construction of the RPG $F[\pi^*]$ from the graph $D[\pi^*]$}: We construct the directed graph $F[\pi^*]$ by exploiting the adjacency relation of the nodes of the dag $D[\pi^*]$, as follows:
\begin{itemize}
\item[(i)\,]
for every vertex $v_i$ of $D[\pi^*]$, $1 \leq i \leq n$, create a node $u_i$ and add it to $V(F[\pi^*])$; create the nodes $u_{n+1}$ and $u_0$ and add them to $V(F[\pi^*])$; note that, the nodes $u_{n+1}$ and $u_0$ correspond to $s$ and $t$, respectively;
\vspace*{0.04in}
\item[(ii)\,]
for every pair of nodes $(u_i, u_{i-1})$ of the set $V(F[\pi^*])$ add the directed edge $(u_i, u_{i-1})$ in $E(F[\pi^*])$, $1 \leq i \leq n+1$;
\vspace*{0.04in}
\item[(iii)\,]
add the directed edge $(u_i, u_j)$ in $E(F[\pi^*])$ if $(v_i, v_j) \in E(D[\pi^*])$, $1 \leq i \leq n+1$, and $v_i$ is the maximum-labeled element of the set $\{v_{i_1}, v_{i_2}, \ldots, v_{i_{indeg(i)}}\}$, where $(v_{i_k}, v_j) \in E(D[\pi^*])$, $1 \leq k \leq indeg(i)$.
\end{itemize}


\vspace*{0.15in} \noindent Algorithm {\tt Encode\_SIP-to-RPG}
\vspace*{-0.04in}
\begin{enumerate}
\item[1.\,]   
Construct a directed acyclic graph (dag) $D[\pi^*]$ on $n$ vertices as follows:\\
\mytab $\circ$ \ $V(D[\pi^*]) = \{v_1, v_2, \ldots, v_n\}$;\\
\mytab $\circ$ \ compute the set $didom(i)$ of each element $i$ in $\pi^*$, $1 \leq i \leq n-1$;\\
\mytab $\circ$ \ for each $j \in didom(i)$, add the edge $(v_i, v_j)$ in $E(D[\pi^*])$;\\
\mytab $\circ$ \ add two dummy vertices $s = v_{n+1}$ and $t = v_0$ in $V(D[\pi^*])$;\\
\mytab $\circ$ \ add $(s, v_i)  \in E(D[\pi^*])$, for every $v_i$ with $indeg(v_i) = 0$;\\
\mytab $\circ$ \ add $(v_i, t)  \in E(D[\pi^*])$, for every $v_i$ with $outdeg(v_i) = 0$;
\vspace*{0.04in}
\item[2.\,]   
For each vertex $v_i \in V(D[\pi^*])$, $1 \leq i \leq n$, do\\
\mytab $\circ$ \ compute the set $P(v_i) = \{v_j \in V(D[\pi^*]) | (v_j, v_i)  \in E(D[\pi^*])\}$;\\
\mytab $\circ$ \ select the maximum-labeled vertex $v_m$ from $P(v_i)$;\\
\mytab $\circ$ \ set $p(v_i) = v_m$;
\vspace*{0.04in}
\item[3.\,]   
Construct a directed graph $F[\pi^*]$ on $n+2$ vertices, as follows:\\
\mytab $\circ$ \ $V(F[\pi^*]) = \{t = u_0, u_1, \ldots, u_n, u_{n+1} = s\}$;\\
\mytab $\circ$ \ for $i$ = $n$ downto 0 do\\
\mytab\mytab \ \ add the edge $(u_{i+1}, u_i)$ in $E(F[\pi^*])$; we call it {\it list pointer};
\vspace*{0.04in}
\item[4.\,]   
For each vertex $u_i \in V(F[\pi^*])$, $1 \leq i \leq n$, do\\
\mytab $\circ$ \ add the edge $(u_i, u_m)$ in $E(F[\pi^*])$ if $v_m = p(v_i)$;\\
\mytab \phantom{$\circ$ } we call it {\it max-didomitation pointer};
\vspace*{0.04in}
\item[5.]   
Return the graph $F[\pi^*]$;
\end{enumerate}

\vspace*{0.04in}
\noindent {\it Time and Space Complexity.} The most time- and space-consuming steps of the algorithm are the construction of the directed graph $D[\pi^*]$ (Step~1) and the computation of the function $p$ for each vertex $v_i \in V(D[\pi^*])$, $1 \leq i \leq n$ (Step~2; recall that $p(v_i)$ equals the maximum-labeled vertex $v_m$ of the set $P(v_i)$ containing all the vertices of $D[\pi^*]$ which didominate vertex $v_i$). On the other hand, the construction of the reducible permutation flow-graph $F[\pi^*]$ (Steps~3 and 4) requires only the list pointers, which can be trivially computed, and the max-didomitation pointers, which can be computed using the function $p$.

Looking at the permutation $\pi^*$, we observe that the element $m$ which corresponds to vertex $v_m$ of $D[\pi^*]$ is the max-indexed element on the left of the element $i$ in $\pi^*$ that is greater than $i$. Thus, the function $p$ can be alternatively computed using the input permutation as follows:

\vspace*{-0.02in}
\begin{enumerate}
\item[(i)]
insert the element $s$ with value $n+1$ into a stack $S$;\\
$top\_S$ is the element on the top of the stack;\\
\vspace*{-0.08in}
\item[(ii)]
for each element $\pi_i \in \pi^*$, $i = 1, 2, \ldots, n$, do the following:\\
\mytab while $top\_S < \pi_i$ do\\
\mytab\mytab remove the $top\_S$ from $S$;\\
\mytab $p(u_i) = top\_S$;\\
\mytab insert $\pi_i$ in stack $S$;
\end{enumerate}
\vspace*{-0.02in}

\noindent Since each element of the input permutation $\pi^*$ is inserted once in the stack $S$ and is compered once with each new element the whole computation of the function $p$ takes $O(n)$ time and space, where $n$ is the length of the permutation $\pi$. Thus, we obtain the following result:

\vspace*{0.04in}
\begin{theorem}\label{theo2.1}
Let $\pi^*$ be a self-inverting permutation of length $n$. The algorithm {\tt Encode\_SIP-to-RPG} for encoding the permutation $\pi^*$ as a reducible permutation flow-graph $F[\pi^*]$ requires $O(n)$ time and space.
\end{theorem}

\begin{figure}[t]
 \hrule \medskip
  \centering
  \includegraphics[scale=0.6]{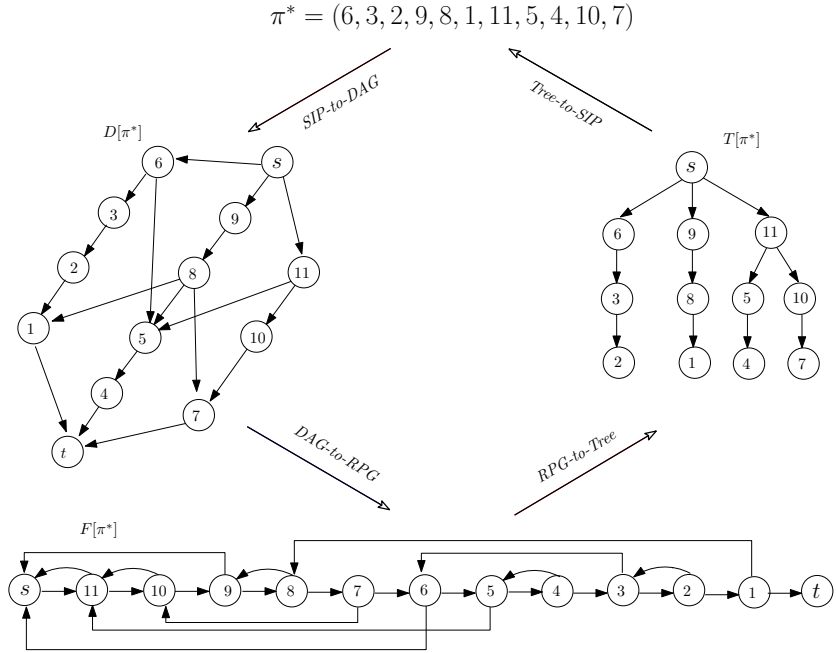}
  \centering
  \caption{\small{The main structures used or constructed by the algorithms {\tt Encode\_SIP-to-RPG} and {\tt Decode\_RPG-to-SIP}; that is, the  self-inverting permutation $\pi^*$, the dag $D[\pi^*]$, the reducible graph $F[\pi^*]$, and the tree $T[\pi^*]$}}
  \label{class} \medskip
 \hrule
\end{figure}

\vskip 0.3in 
\subsection{Algorithm Decode\_RPG-to-SIP}

The algorithm {\tt Encode\_SIP-to-RPG} produces reducible permutation flow-graph $F[\pi^*]$ in which it encodes a self-inverting permutation $\pi^*$. Thus, we are interested in designing an efficient and easily implemented algorithm for extracting the permutation $\pi^*$ from the graph $F[\pi^*]$.

Next, we present such a decoding algorithm, we call it {\tt Decode\_RPG-to-SIP}, which is efficient: it takes time and space linear, i.e., $O(n)$, in the size of the flow-graph $F[\pi^*]$, and easily implemented: the only operations used by the algorithm are edge modifications on $F[\pi^*]$ and DFS-search on trees.

The algorithm takes as input a reducible permutation flow-graph $F[\pi^*]$ on $n+2$ nodes and produces a self-inverting permutation $\pi^*$ of length $n$; it works as follows:


\vspace*{0.15in} \noindent Algorithm {\tt Decode\_RPG-to-SIP}
\vspace*{-0.04in}
\begin{enumerate}
\item[1.\,]   
Delete the directed edges $(v_{i+1}, v_i)$ from the edge set $E(F[\pi^*])$, $1 \leq i \leq n$,
and the node $t = v_0$ from $V(F[\pi^*])$;
\vspace*{0.04in}
\item[2.\,]   
Flip all the remaining directed edges of the graph $F[\pi^*]$; let $T[\pi^*]$ be the resulting tree
and let $s=v_0, v_1, v_2, \ldots, v_n$ be the nodes of $T[\pi^*]$;
\vspace*{0.04in}
\item[3.\,]   
Perform DFS-search on tree $T[\pi^*]$ starting at node $s$ by always proceeding to the minimum-labeled child node
and compute the DFS discovery time $d[v]$ of each node $v$ of $T[\pi^*]$;
\vspace*{0.04in}
\item[4.\,]   
Order the nodes $s=v_0, v_1, v_2, \ldots, v_n$ of the tree $T[\pi^*]$ by their DFS discovery time
$d[]$ and let $\pi=(v'_0, v'_1, v'_2, \ldots, v'_n)$ be the resulting order, where $d[v'_i]<d[v'_j]$ for $i<j$, $0 \leq i, j \leq n$;
\vspace*{0.04in}
\item[5.\,]   
Delete node $s$ from the order $\pi$;
\vspace*{0.04in}
\item[6.\,]   
Return $\pi^* = \pi$;
\end{enumerate}

\vspace*{0.04in}
\noindent {\it Time and Space Complexity.} The size of the reducible permutation graph $F[\pi^*]$ constructed by the algorithm {\tt Encode\_SIP-to-RPG} is $O(n)$, where $n$ is the length of the permutation $\pi^*$, and thus the size of the resulting tree $T[\pi^*]$ is also $O(n)$. It is well known that the DFS-search on the tree $T[\pi^*]$ takes time linear in the size of $T[\pi^*]$. Thus, the decoding algorithm is executed in $O(n)$ time using $O(n)$ space. Thus, the following theorem holds:

\vspace*{0.04in}
\begin{theorem}\label{theo1.2}
Let $F[\pi^*]$ be a reducible permutation flow-graph of size $O(n)$ produced by the algorithm {\tt Encode\_SIP-to-RPG}. The algorithm {\tt Decode\_RPG-to-SIP} decodes the flow-graph $F[\pi^*]$ in $O(n)$ time and space.
\end{theorem}

\section{Properties and Attacks}
In this section, we analyze the structures of the two main components of our proposed codec, that is, the self-inverting permutation $\pi^*$ produced by the algorithm {\tt Encode\_W-to-SIP} and the reducible permutation graph $F[\pi^*]$ produced by the algorithm {\tt Encode\_SIP-to-RPG}, and discuss their properties with respect to resilience to attacks.

\vskip 0.3in 
\subsection{Properties of permutation $\pi^*$}

Collberg et al. \cite{CT99,CKCT03} describe several techniques for encoding watermark integers in graph structures. Based on the fact that there is a one-to-one correspondence between self-inverting permutations and isomorphism classes of RPGs, Collberg et al. \cite{CKCT03} proposed a polynomial algorithm for encoding any integer $w$ as the RPG corresponding to the $w$th self-inverting permutation $\pi$ in this correspondence.
This encoding exploits only the fact that a self-inverting permutation is its own inverse; it does not incorporate any other structural property.

In our codec system proposed in this paper an integer $w$ is encoded as self-inverting permutation $\pi^*$ using a particular construction technique which captures into $\pi^*$ important structural properties. These properties unable us to identify any single change (in some cases, multiple changes) made by an attacker to $\pi^*$.

The main structural properties of our self-inverting permutation $\pi^*$ produced by the algorithm {\tt Encode\_W-to-SIP} can be summarized into the following three categories:

\begin{enumerate}
\item[$\bullet$\,] {\bf Length property}: By construction the self-inverting permutation $\pi^*$ has always odd length. Thus, any single node-modification, i.e., adding an element in $\pi^*$ or deleting an element from $\pi^*$, can be easily identified;

\vspace*{0.04in}
\item[$\bullet$\,] {\bf Bitonic property}: Algorithm {\tt Decode\_SIP-to-W} decodes the self-inverting permutation $\pi^*$ to obtain the encoded integer $w$. During the decoding process two sequences are constructed, that is, the increasing subsequence $X$ and the decreasing subsequence $Y$ (see Step~4), which incorporate the bitonic property of the encoding process. If the permutation $\pi^*$ has not been produced by our encoding algorithm {\tt Encode\_W-to-SIP} then subsequence $Y$ may not be increasing. Thus, an appropriate change to SIP $\pi^*$ that keeps the SIP property may be identified by checking the subsequence $Y$;

\vspace*{0.04in}
\item[$\bullet$\,] {\bf Block property}: The algorithm {\tt Encode\_W-to-SIP} takes the binary representation of the integer $w$ and constructs the number $B'$ (see Step~2). The binary representation of $B'$ consists of three parts (or, blocks): (i) the first part contains the first $n$ bits with 0s values, (ii) the second part contains the next $n$ bits which forms the binary representation of the integer $w$, and (iii) the third part of length one contains a bit 1. This property is encapsulated in the structure of $\pi^*$ in such a way that during the decoding process the binary sequence $B'$ constructed in Step~6 of the decoding algorithm {\tt Decode\_SIP-to-W} is identical to the sequence $B'$ constructed by the encoding algorithm {\tt Encode\_W-to-SIP}. If an attacker make appropriate changes to SIP $\pi^*$ so that the resulting permutation $\pi^*$ still has the SIP property, then the first block of the binary sequence $B'$ may contain one or more 1s or the third block may be 0.
\end{enumerate}

\section{Properties of the Flow-graph $F[\pi^*]$}

We next describe the main properties of our codec system $({\tt encode}, {\tt decode})_{F[\pi^*]}$; we mainly focus on the properties of the reducible permutation graph $F[\pi^*]$ with respect to graph-based software watermarking attacks.

\vspace*{0.15in}
\noindent {\it Components of the graph $F[\pi^*]$}: The reducible permutation graph $F[\pi^*]$ consists of the following three components:
\begin{enumerate}
\item[(1)] {\bf A header node}: it is a root node with outdegree one from which
every other node in the graph $F[\pi^*]$ is reachable. Note that, every control flow-graph has
such a node. In the graph $F[\pi^*]$ the header node is denoted by $s$;
\vspace*{0.04in}
\item[(2)] {\bf A footer node}: it is a node with outdegree zero that is reachable from every other
node of the graph. Every control flow-graph has such a node, representing
the method exit. In the graph $F[\pi^*]$ the footer node is denoted by $t$;
\vspace*{0.04in}
\item[(3)] {\bf The body}: it consists of $n$ nodes $u_1, u_2, \ldots, u_n$ each with outdegree two. In particular, each node $u_i$ $(1 \leq i \leq n)$ has exactly two outpointers: one points to node $u_{i-1}$, which we call {\it list pointer}, and the other points to node $u_m$, which we call {\it max-didomination pointer}, where $m>i$; note that, $u_m > u_i > u_{i-1}$.
\end{enumerate}

\noindent {\it Structural Properties}: In graph-based encoding algorithms, the watermark $w$ is encoded into some special
kind of graphs $G$. Generally, the watermark graph $G$ should not differ from the graph data structures
built by real programs. Important conditions are that the maximum outdegree of $G$
should not exceed two or three, and that the graph $G$ have a unique root node so the
program can reach other nodes from the root node. Moreover, $G$ should be resilient to
attacks against edge and/or node modifications. Finally, $G$ should be efficiently constructed.

The proposed reducible permutation graph $F[\pi^*]$ and a corresponding codec $({\tt encode}, {\tt decode})_{F[\pi^*]}$ have all the above properties; in particular, the graph $F[\pi^*]$ and the corresponding codec have the following properties:
\begin{enumerate}
\item[$\bullet$\,] {\bf Appropriate graph types}: The graph $F[\pi^*]$ is directed on $n+2$ nodes with outdegree exactly two; that is, it has low max-outdegree, and, thus, it matches real program graphs;
\vspace*{0.04in}
\item[$\bullet$\,] {\bf High resiliency}: Since each node in the reducible permutation graph $F[\pi^*]$ has exactly one list outpointer and exactly one max-didom outpointer, any single edge modification, i.e., edge-flip, edge-addition, or edge-deletion, will violate the outpointer condition of some nodes, and thus the modified edge can be easily identified and corrected. Thus, the graph $F[\pi^*]$ unable us to correct single edge changes;
\vspace*{0.04in}
\item[$\bullet$\,] {\bf Small size}: The size $|P_w| - |P|$ of the embedded watermark is small;
\vspace*{0.04in}
\item[$\bullet$\,] {\bf Efficient codecs}: The codec $({\tt encode}, {\tt decode})_{F[\pi^*]}$ has low time and space complexity; indeed, we have showed (see Theorem~4 and Theorem~5) that the encoding algorithm {\tt Encode\_SIP-to-RPG} requires $O(n)$ time and space, where $n$ is the size of the input permutation $\pi^*$, while the decoding algorithm {\tt Decode\_RPG-to-SIP} decodes the flowgraph $F[\pi^*]$ in $O(n)$ time and space.
\end{enumerate}

\noindent It is worth noting that our encoding and decoding algorithms use basic data structures and basic operations, and, thus, they can be easily implemented.

\vspace*{0.15in}
\noindent {\it Unique Hamiltonian Path}: It is well-known that any acyclic digraph $G$ has at most one Hamiltonian path (HP); $G$ has one HP if the subgraphs $G_0, G_1, \ldots, G_n$ have only one node with indegree zero, where $G_0=G$ and $G_i = G \backslash \{v_1, v_2, \ldots, v_i\}$, $1 \leq i \leq n-1$ (recall that $n$ denotes the number of nodes in $G$). Furthermore, it has been shown that any reducible flow-graph has at most one Hamiltonian path \cite{CKCT03}.

We next show that the reducible permutation graph $F[\pi^*]$ produced by the algorithm {\tt Encode\_SIP-to-RPG} has always a unique Hamiltonian path, denoted by HP$(F[\pi^*])$, and this Hamiltonian path can be found in $O(n)$ time, where $n$ is the number of nodes of $F[\pi^*]$. The following algorithm, which we call {\tt Unique\_HP}, takes as input a flow-graph $F[\pi^*]$ on $n$ nodes and produces its unique Hamiltonian path HP$(F[\pi^*])$.


\vspace*{0.15in} \noindent Algorithm {\tt Unique\_HP}
\vspace*{-0.04in}
\begin{enumerate}
\item[1.\,]   
Find the node $u_0$ of the graph $F[\pi^*]$ with outdegree one;
\vspace*{0.04in}
\item[2.\,]   
Perform DFS-search on graph $F[\pi^*]$
starting at node $u_0$ and compute the DFS discovery time $d[u]$ of each node $u$ of $F[\pi^*]$;
\vspace*{0.04in}
\item[3.\,]   
Order the nodes $u_0, u_1, \ldots, u_{n+1}$ of the graph $F[\pi^*]$ by their DFS discovery time $d[]$ and let
HP$(F[\pi^*])=(u'_0, u'_1, \ldots, u'_{n+1})$ be the resulting order, where $d[u'_i]<d[u'_j]$ for $i<j$, $0 \leq i, j \leq n+1$;
\vspace*{0.04in}
\item[4.\,]   
Return HP$(F[\pi^*])$;
\end{enumerate}

\vspace*{0.02in}
Since the graph $F[\pi^*]$ contains $n$ nodes and $m=O(n)$ edges, both finding the node of $F[\pi^*]$ with outdegree one and performing DFS-search on $F[\pi^*]$ take $O(n)$ time and require $O(n)$ space. Thus, we have the following result.

\vspace*{0.04in}
\begin{theorem}\label{theo5.1}
Let $F[\pi^*]$ be a reducible permutation graph of size $O(n)$ produced by the algorithm {\tt Encode\_SIP-to-RPG}. The algorithm {\tt Unique\_HP} correctly computes the unique Hamiltonian path of $F[\pi^*]$ in $O(n)$ time and space.
\end{theorem}

\vspace*{0.02in}

By construction, our reducible flow-graph $F[\pi^*]$ is a labeled graph;
graphs in which labels (which are most commonly numbers) are assigned to nodes are called labeled graphs, while graphs in which individual nodes have no distinct identifications are called unlabeled graphs (unless indicated otherwise by context, the term ``labeled" graph generally refers to a node labeled graph). Indeed, the labels of $F[\pi^*]$ are numbers of the set $\{0, 1, \ldots, n+1\}$, where the label $n+1$ is assigned to header node $s=u_{n+1}$, the label $0$ is assigned to footer node $t=u_{0}$, and the label $n-i$ is assigned to the $i$th body node $u_{n+1-i}$, $1 \leq i \leq n$.

Let $F'[\pi^*]$ be the graph which results after making some label modifications on the flow-graph $F[\pi^*]$; a label modification attacker may be performs swapping of the labels of two nodes, altering the value of the label of a node, or even removing all the labels of the graph $F[\pi^*]$ resulting an unlabeled graph. Since the extraction of the watermark $w$ relies on the labels of the flow-graph $F[\pi^*]$ (see algorithm {\tt Decode\_RPG-to-SIP}), it follows that our codec system $({\tt encode}, {\tt decode})_{F[\pi^*]}$ is susceptible to node modification attacks.

Thus, we are interested in finding a way to extract the watermark $w$ efficiently from $F[\pi^*]$ without relying on its labels; for example, to extract $w$ efficiently from the graph $F'[\pi^*]$. We show that, after any node modification attack on graph $F[\pi^*]$, we can efficiently reassign the initial labels to nodes of $F[\pi^*]$ using the structure of the unique Hamiltonian path HP$(F[\pi^*])$. More precisely, given the graph $F'[\pi^*]$ we can construct the flow-graph $F[\pi^*]$ in $O(n)$ time and space. In addition, if $F'[\pi^*]$ is the unlabeled graph of the flow-graph $F[\pi^*]$ we can also construct the graph $F[\pi^*]$ in $O(n)$ time and space.

The above properties imply that we are able to extract a watermark $w$ in linear time from a node modified or unlabeled flow-graph $F[\pi^*]$; this can be simply done by assigning labels to nodes of $F[\pi^*]$ just prior the use of the decoding algorithm {\tt Decode\_RPG-to-SIP}. Thus, we obtain the following result.

\vspace*{0.04in}
\begin{lemma}\label{lemm5.1}
Let $F[\pi^*]$ be a reducible permutation graph of size $O(n)$ produced by the algorithm {\tt Encode\_SIP-to-RPG} and let $F'[\pi^*]$ be the graph resulting from $F[\pi^*]$ after modifying or deleting the node-labels of $F[\pi^*]$. Given $F'[\pi^*]$, the flow-graph $F[\pi^*]$ can be constructed in $O(n)$ time and space.
\end{lemma}

\section{Concluding Remarks}

In this paper we extended the class of error correcting graphs by proposing efficient and easy to implement graph encodings. In particular, we proposed an efficient and easily implemented codec system for encoding watermark numbers as graph structures.

More precisely, we first presented the algorithm {\tt Encode\_W-to-SIP} which encodes an integer $w$ as SIP (self-inverting permutation) $\pi^*$ in $O(n)$ time and space, where $n$ is the number of bits in the binary representation of $w$, and the corresponding decoding algorithm {\tt Decode\_SIP-to-W} which extracts the watermark number $w$ from the SIP $\pi^*$ also in $O(n)$ time and space.

We next presented the algorithm {\tt Encode\_SIP-to-RPG} which encodes the SIP $\pi^*$ as a reducible flow-graph $F[\pi^*]$ in $O(n)$ time and space by exploiting didomination relations on the elements of $\pi^*$, and the corresponding decoding algorithm {\tt Decode\_RPG-to-SIP} which extracts the SIP $\pi^*$ from the graph $F[\pi^*]$ in $O(n)$ time and space by converting first the graph $F[\pi^*]$ into a directed tree $T[\pi^*]$ and then applying DFS-search on $T[\pi^*]$.

The main features of our proposed encoding and decoding algorithms can be summarized as follows:

\vspace*{-0.06in}
\begin{enumerate}
\item[$\bullet$\,] {\bf Algorithms {\tt Encode\_W-to-SIP} and {\tt Decode\_SIP-to-W}}: use basic data structures; apply elementary operations on sequences; have low time and space complexity; have an easy implementation;
\vspace*{0.04in}
\item[$\bullet$\,] {\bf Algorithms {\tt Encode\_SIP-to-RPG} and {\tt Decode\_RPG-to-SIP}}: use domination relations on permutations; construct dags and lists; use DFS-search on directed trees; have low time and space complexity; have an easy implementation;
\end{enumerate}

\vspace*{-0.06in}
\noindent An interesting property of our encoding approach is that of enabling us to encode the integer $w = b_1b_2 \cdots b_n$ as self-inverting permutation $\pi^*$ of any length; indeed, $\pi^*$ can be constructed over the set $N_{n'} = \{1, 2, \ldots, n'\}$, where the smallest value of $n'$ is $O(\log n)$.

It is worth noting that the two main components of our proposed codec system, i.e., the self-inverting permutation $\pi^*$ and the reducible permutation graph $F[\pi^*]$, incorporate important structural properties, due to the bitonic property encapsulated in $\pi^*$ and the reducible property of  $F[\pi^*]$, which make our codec system resilient to attacks. In particular, these properties unable us to identify any single change (in some cases, multiple changes) made by an attacker to $\pi^*$ and $F[\pi^*]$.

Thus, in light of our two codec components $\pi^*$ and $F[\pi^*]$ proposed in this paper it would be very interesting to come up with new efficient codec algorithms and structures having ``better" properties with respect to resilience to attacks; we leave it as an open question.
An interesting question with practical value is whether the class of reducible permutation graphs can be extended so that it includes other classes of graphs with structural properties capable to efficiently encode watermark numbers.

Another interesting question with practical value is whether we can produce more than one reducible flow-graphs $F_1[\pi^*], F_2[\pi^*], \ldots, F_n[\pi^*]$ which encode the same self-inverting permutation $\pi^*$; we also leave it as an open question (see \cite{CN11} for cographs).

Finally, we leave as an open problem the evaluation of our codec algorithms and structures in a simulation environment in order to obtain detailed information about their practical behaviour. For future investigation, we also leave as an open problem the analysis of our codec algorithms under other software watermarking measurements.

\end{document}